\documentclass[aps,prl,twocolumn,showpacs,amsmath,amssymb,unsortedaddress,floatfix]{revtex4-1}
\usepackage{graphicx}
\usepackage{tabularx}
\usepackage{color}
\usepackage{natbib}

\newcommand{\Ang}{\,\mathrm{\AA}}

\def\eV{\,\textrm{eV}}
\def\etal{\textit{et al.}}
\newcommand{\fom}{\,\mathrm{f\Omega m^2}}

\begin{document}

\title{Electron Transport Through Ag-Silicene-Ag Junctions}

\author{Yun-Peng Wang,$^{1,2} $ J. N. Fry,$^1$ and Hai-Ping Cheng$^{1,2}$ } 
\email[Corr. author: Hai-Ping Cheng,  ]{cheng@qtp.ufl.edu}
\affiliation{$^1$Department of Physics, University of Florida,  Gainesville, Florida 32611, USA \\
             $^2$Quantum Theory Project, University of Florida, Gainesville, Florida 32611, USA}

\begin{abstract}

For several years the electronic structure properties of the novel two-dimensional system silicene have been studied extensively.
Electron transport across metal-silicence junctions, however, remains relatively unexplored.
To address this issue, we developed and implemented a theoretical framework that utilizes the tight-binding Fisher-Lee relation to span
non-equilibrium Green's function (NEGF) techniques, the scattering method, and semiclassical Boltzmann transport theory. 
Within this hybrid quantum-classical, two-scale framework, 
we calculated transmission and reflection coefficients of monolayer and bilayer Ag-silicene-Ag junctions
using the NEGF method in conjunction with density functional theory;
derived and calculated the group velocities; 
and computed resistance using the semi-classical Boltzmann equation. 
We found that resistances of these junctions are 
$\sim${}$ 0.08 \fom$ for monolayer silicene junctions and $\sim${}$ 0.3 \fom$ for bilayer ones, 
factors of $\sim$8 and $\sim$2, respectively, smaller than Sharvin resistances estimated via the Landauer formalism. 
\end{abstract}

\maketitle

%\section{introduction}

The novel two-dimensional system silicene has received recent extensive attention.
Silicene has a buckled honeycomb lattice structure similar to graphene.
Band structure calculations predicted a linear dispersions near the Fermi level \cite{PhysRevLett.102.236804},
so low-energy excitations in silicene can be described using the Dirac equation with zero rest mass.
One advantage of silicence over graphene is its tunable electronic structure, 
because of the buckling in the lattice \cite{ncomms.4.1500}, 
by applying an electric field normal to the silicene sheet.
Another advantage compared to graphene is its compatibility with modern silicon-based electronics technology.
%In fact silicene is basically a single (111) plane of bulk Si (lattice constant a=5.43 $\Ang$).
%The (111) plane of bulk Si has a honeycomb atomic structure with lattice constant of 3.84 $\Ang$
%and the buckling distance of 0.39 $\Ang$,
%which are very close to those in low-buckled silicene (3.83 and 0.44 $\Ang$) \cite{PhysRevLett.102.236804}.
%Recently hydrogen-terminated Si was proposed theoretically to be a good substrate for silicene \cite{arxiv.1211.3495}.
However, interactions between Si(111) layers are extremely strong, and 
the mechanical exfoliation method used to produce graphene from graphite cannot be used to obtain silicene sheets. 
The only successful approach so far has been to deposit Si atoms onto substrates with six-fold rotation symmetry,
including Ag(111), $\mathrm{ZrB}_2$, and Ir(111). 
Unfortunately, all of these substrates are metallic, and 
there has been no report of success transferring silicene from one substrate onto another,
in particular onto an insulating substrate. 
As a result, although a number of theoretical investigations have been performed to address 
transport properties, including maganetoresistance, of standalone silicene nano-ribbons \cite{APL.100.233122}, 
it might be difficult to study the in-plane conductance of silicene directly in experiments.

In this Letter, we report a theoretical investigation
on electron transport through Ag-silicene-Ag junctions
and propose a future experimental study focusing on 
current-perpendicular-to-plane (CPP) measurements.
The CPP transport properties of silicene
can be readily measured in present experiments.
Silicene monolayers and bilayers have been successfully synthesized by depositing Si atoms onto Ag(111) surfaces 
\cite{PhysRevLett.108.155501,PhysRevLett.109.056804},
based on which we designed a Ag(111)-silicene-Ag junction by depositing another Ag(111) lead on top of silicene.
In this way the CPP transport properties of silicene can be studied.
Ag is chosen in this work because it is an ideal material for a silicene substrate, 
with relatively small lattice mismatch and low tendency to form Si-Ag alloy.

Based on first-principles density functional theory (DFT), we thoroughly investigated the interface
structures of monolayer- and bilayer-silicene junctions, from which we used
non-equilibrium Green's function (NEGF) 
\cite{Datta1995,PhysRevB.63.245407,Chem.Phys.281.151}
techniques to compute 
transmissions of Ag-silicene-Ag junctions.
We found that transmissions of these junctions are
about 70\% that of perfect Ag junctions for monolayer silicene junctions
and about 35\% for bilayer junctions.
The Landauer-Buttiker equation \cite{PhysRevB.23.6851,PhysRevB.31.6207}
developed for narrow conduction channels (or 1D systems) can be used to estimate the Sharvin resistance,
but the Sharvin resistance is no longer an experimental observable quantity  for 2D junctions 
where the transmission coefficient per channel is a fraction of unity.  
To overcome this difficulty, we developed a two-scale, hybrid quantum-classical framework that employs 
first-principles NEGF techniques and a tight-binding (TB) Fisher-Lee relation \cite{wimmer} to calculate transmission coefficients
and the  semi-classical Boltzmann equation method \cite{JS.13.221} to
calculate conductances of junctions. 
The two ingredients to this approach are the derivation of an expression for the group velocity of Bloch state from the NEGFs 
and the implementation of the scheme in a major computation package.

%The rest of this paper is organized as follows:
%In Section II, interfacial atomic structures of monolayer- and bilayer-silicene junctions
%optimized using DFT method and transmissions of junctions using NEGF method are presented.
%In Section III, the two-scale, hybird quantum-classical approach is built
%,and the generalized Fisher-Lee relation and expression for group velocity are derived.
%The resistances of Ag-silicene-Ag junctions were calculated via semiclassical Boltzmann equation.
%Section IV is a summary on the theoretical framework developed in this paper.
%\section{First-principles calculations}
%Atomic structures of Ag-silicene-Ag junctions were fully optimized using 
%{\it ab initio} (DFT) method,
%based on which transmission through junctions were studied using NEGF method.
%\subsection{Interface Structures}

\begin{figure}[h]
\begin{center}
\includegraphics[width=0.6\linewidth]{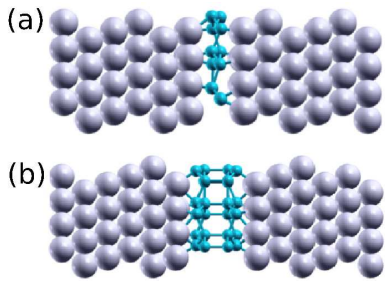}
\caption{
\label{fig:atoms}
(Color online)
Atomic structures of Ag-silicene-Ag junctions with (a) monolayer silicene and (b) bilayer silicene in the ``AA'' stacking configuration (see text).
Big(small) spheres represent Ag(Si) atoms.
}
\end{center}
\end{figure}

Atomic structures of Ag-silicene-Ag junctions were fully optimized using 
the \textit{ab initio} DFT method.
The lattice constant of the Ag(111) surface ($2.95\Ang$) is about 0.75 times that of the low-buckled silicene lattice ($3.83\Ang$),
so we chose a supercell containing $4\times4$ Ag(111) unit cells and $3\times3$ silicene unit cells, 
with the lattice vectors of the two subsystems parallel.
The atomic structures of Ag-silicene-Ag junctions with monolayer and bilayer silicene are shown in Fig.~\ref{fig:atoms},
constructed according to the atomic structure of the silicene-Ag system proposed in Ref.~\onlinecite{PhysRevLett.108.155501}.
%using scanning tunnelling microscopy and first-principle calculations.
Note that several different atomic arrangements have been observed in the silicene-Ag system
(see Ref.~\onlinecite{0953-8984-24-31-314211} for a review); 
the atomic structure chosen here is the most common one observed in experiments.
Atomic structures of junctions were fully optimized.
During structural optimization, Si atoms in silicene layers and Ag atoms in the nearest and the next-nearest layers
with respect to the silicene layers were allowed to relax.
The total energy and forces on atoms were calculated using 
DFT with the the Perdew-Burke-Ernzerhof (PBE) parametrization \cite{PhysRevLett.77.3865}
of the generalized gradient approximation, 
as implemented in the Vienna {\it Ab initio} Simulation Package 
{\tt VASP} \cite{PhysRevB.59.1758}.
The PBE functional has been proven to produce the same atomic structure as 
vdW-DF functionals \cite{PhysRevLett.110.076801}.
Periodic boundary conditions were used in the $z$-direction 
(perpendicular to the plane of the silicene sheet).
Forces on atoms were converged to be smaller than  $ 0.01 \eV/\textrm{\AA} $. 
For several inter-electrode distances the geometry was optimized as described above, 
and the optimal inter-electrode distance was taken as the one with the lowest total energy. 

There are three different configurations for each kind of junction 
due to different stacking relations between the two Ag(111) layers
adjacent to silicene sheet in the left and right leads. 
Bulk Ag has a face-centered cubic lattice, and the atomic layer along the [111] direction has an ``ABC'' stacking.
The Ag(111) layer adjacent to silicene on the left side was always labeled as ``A'', 
and the first Ag layer on the right lead can then be ``A'', ``B'', or ``C''. 
All of three configurations were considered in our calculations.

\begin{figure}[h]
\begin{center}
\includegraphics[width=0.8\linewidth]{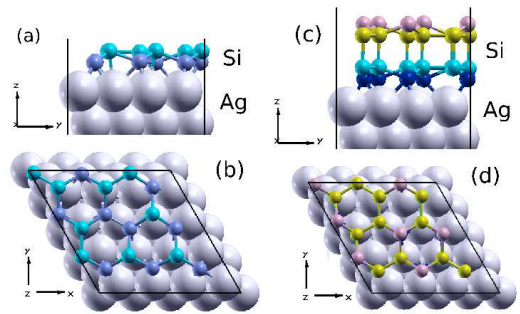}
\caption{
\label{fig:cartoon}
(Color online)
(a) Side view and (b) top view of the interface structure of Ag-monolayer silicene-Ag junctions; 
(c) side view and (d) top view of the interface structure of Ag-bilayer silicene-Ag junctions.
Only Ag atoms (big spheres) in the left lead are shown for clarity, 
and Si atoms (small spheres) with different $z$-positions are labelled in different colors.
The interface structure is independent of the stacking of Ag(111) layer in the right lead (see text).
}
\end{center}
\end{figure}

In the fully optimized structure of monolayer silicene junctions,
half of the Si atoms move closer to the left lead 
and the other half closer to the right lead, 
which is different from the atomic structure of silicene on a single Ag(111) surface \cite{PhysRevLett.108.155501}.
The three different  configurations with the first Ag layer in the right lead in an ``A'', ``B'', or ``C'' stacking
have the same interface structure, in particular the same lattice buckling pattern of silicene.
The interface structure is shown in Fig.~\ref{fig:cartoon}(a) and (b), where only the left lead and silicene atoms 
are plotted and Si atoms with different $z$-positions are labelled in different colors.
The structures of bilayer silicene junctions were also optimized, and 
their interface structure is shown in Fig.~\ref{fig:cartoon}(c) and (d).
Silicon atoms in the top silicene layer (Fig.~\ref{fig:cartoon}(c)) 
can be either at hollow sites or at sites atop atoms with respect to the bottom silicene layer.
To determine the ground state, 
optimizations were performed using these two possible structures as starting point
and we found that the atop structure is the ground state, 
which is consistent with the atomic structure of bulk Si.
The optimized buckling patterns are again independent of the stacking of the right lead.

%\subsection{{\it Ab initio} NEGF Calculations}

\begin{figure}[h]
\begin{center}
\includegraphics[width=\linewidth]{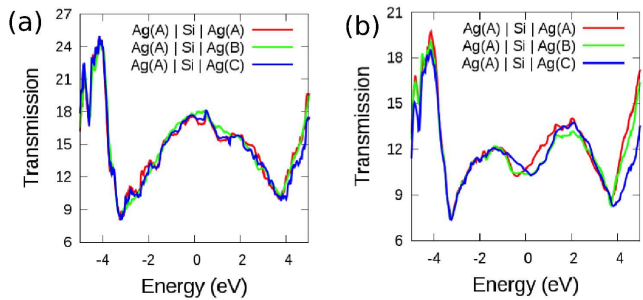}
\caption{
\label{fig:transmission}
(Color online)
Transmissions as a function of energy for junctions with (a) monolayer silicene and
(b) bilayer silicene.
For each junction, there are three different configurations given by different stacking relations (see text).
The Fermi energy is set to zero.
}
\end{center}
\end{figure}

With the optimized interface structures of junctions,
we used the DFT/PBE-based NEGF method to 
compute self-consistently the Green's function in the scattering region.
The scattering region was chosen to include the silicene layer(s) and the three adjacent Ag layers in each lead.
Strictly localized atomic orbitals were used to expand the Hamiltonian and Green's function.
Norm-conserving pseudopotentials \cite{PhysRevB.43.1993} were used to describe interactions between valence electrons
($3s^23p^2$ for Si and $4d^{10}5s^1$ for Ag) and core electrons.
The NEGF calculations were performed using the \texttt{TRANSIESTA} code \cite{PhysRevB.65.165401}.
Translational symmetry in the $x$-$y$ plane was exploited  
(the direction of transport was chosen as the $z$-direction), 
and a sufficient number of $\vec {k}_\| $-points ($\vec{k}_\| \perp \hat{z}$)  
were used both in calculating self-consistent Green's functions and transmissions.
Transmission at a given energy $E$ was calculated as
\begin{equation}
\label{equ:1}
T(E) = \frac{1}{N_\|} \sum_{\vec{k}_\|} T(\vec{k}_\|;E)
\end{equation}
where $N_\|$ is the number of $\vec{k}_\|$'s and 
$T(\vec{k}_\|;E)$ is the transmission of a $\vec{k}_\|$-point  at energy $E$.
Note that Eq.~(\ref{equ:1}) can be applied at any energy, but the transmission at Fermi energy of course is the most interesting.
Calculated transmissions as a function of energy for different junctions are shown in Fig.~\ref{fig:transmission}.

For junctions with a silicene monolayer, the three different configurations 
have transmission curves very close to each other (Fig.~\ref{fig:transmission}(a)).
These transmission curves show broad peaks around Fermi energy with $T(E_F)=17.5$, 
and decrease as energy increases up to $ 4 \eV $ or down to $-3.5 \eV $.
Our analysis shows that these transmission curves are not rooted in monolayer silicene,
which is semi-metallic near the Fermi energy \cite{PhysRevLett.102.236804}.
In order to understand the shape of these transmission curves,
we calculated  a perfect Ag(111)-Ag-Ag(111) junction with all Ag atoms on their bulk lattice positions, 
and found that its transmission curve has a shape very similar to the Ag-silicene-Ag junction 
(Fig.~\ref{fig:transmission}(a)) with $T(E_F)=25.6 $.
In addition, the calculated projected DOS of {\it fcc} Ag leads shows
that the $p$-DOS has a shape similar to the transmission curves of monolayer silicene junctions,
indicating that $p$-electrons in the Ag leads dominate the features of electron transmission through junctions.
At the Fermi energy, transmissions of Ag-silicene-Ag junctions are as large as 70\% of the perfect Ag junction,
%, so that the transport in Ag-silicene-Ag junction is dominated by Ag leads.
indicating low scattering barriers.
%This junction is  and has an averaged transmission per channel of $0.7$.
For bilayer silicene junctions, interesting dips emerge in the transmission curves near the Fermi level,
while the rest of the curves have a shape similar to the monolayer silicene junction (see Fig.~\ref{fig:transmission}(b)). 
The transmission per channel decreases to be 35\% of the Ag-Ag-Ag junction.

\begin{figure}[t]
\begin{center}
\includegraphics[width= \linewidth]{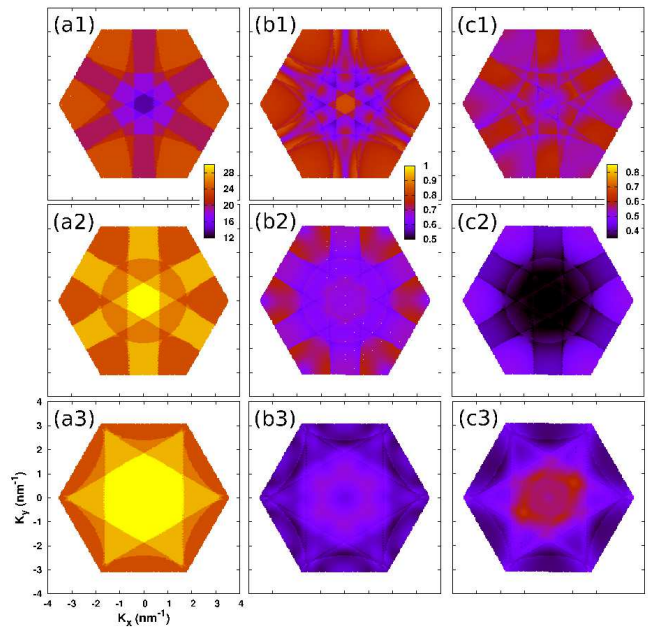}
\caption{
\label{fig:3-energy}
(Color online) 
(a) The numbers of Bloch states in the $4\times 4$ Ag(111) leads, 
(b) normalized transmissions of monolayer silicene junctions, and (c) of bilayer silicene junctions 
with ``AA'' stacking configuration at (1) $-1 \eV$ , (2) $0\eV $, and (3) $1\eV$  with respect to the Fermi energy.
}
\end{center}
\end{figure}

In order to understand the electron transmissions of these junctions in details,
we calculated the normalized transmission $T^n(\vec{k}_{\|})$
which is defined as the ratio between the transmission at $\vec{k}_{\|}$
and the number of Bloch states $ n_B$ in the leads at the same $k$-point,
\begin{equation}
T^n(\vec{k}_{\|})=T(\vec{k}_{\|})/n_B(\vec{k}_{\|})
\end{equation}
The number of Bloch states in leads is equal to the transmission of perfect Ag-Ag-Ag junction
at the corresponding $\vec{k}_{\|}$.
By definition, the normalized transmission is no larger than 1.
In the first column of Fig.~\ref{fig:3-energy},
the numbers of Bloch states in leads for each $\vec{k}_{\|}$ at $-1\eV$ (top), $0\eV$ (middle), and $1 \eV$ (bottom)
are plotted, exhibiting the six-fold rotational symmetry.
The $\vec{k}_\|$-resolved normalized transmissions of monolayer and bilayer silicene junctions with the ``AA'' stacking configuration
at $-1 \eV $ (top),  $0 \eV $ (middle), and $1\eV$ (bottom) are shown in the middle and right-hand column of Fig.~\ref{fig:3-energy}, respectively.
For the monolayer silicene junction, the normalized transmissions at the Fermi energy
are smaller than those at $-1 \eV $, but larger than those at $1 \eV$.
For bilayer silicene junctions (right-hand column in Fig.~\ref{fig:3-energy}),
normalized transmissions at the Fermi energy are smaller than those at  $-1 \eV$  and at $1\eV $.
In particular, the normalized transmissions at the Fermi energy are suppressed to $\sim0.35$ around the $\Gamma$-point,
which results in the dip of the transmission curves around the Fermi energy (Fig.~\ref{fig:transmission}(b)).

%\section{resistances of junctions}
%\subsection{Theory}

Now we need to find an appropriate theory to compute the resistance.
The Landauer-Buttiker formula \cite{PhysRevB.23.6851,PhysRevB.31.6207}, 
which has been used to study nanowires, molecular junctions, break junctions and tunnel junctions that are either narrow channels
or have very small normalized transmissions, 
offers a direct way to calculate resistance from the transmission.
However, the resistance calculated by this approach corresponds to the Sharvin resistance and introduces a significant error
for electron transport  in two- or three-dimensional space where normalized transmissions are a fraction of unity, 
exactly the situation for the present systems.
In these cases, an adequate method for computing conductance % of Ag-silicence-Ag junstion
is the semi-classical Boltzmann (SCB) transport theory, as formulated by Butler \etal{}  in Ref.~\onlinecite{JS.13.221}.
%In their work, a layer-by-layer approach based on scattering theory
%was used to compute the total transmission and reflection coeeficients of the junctions,
%and the group velocity.
In their theory, electrons in both leads follow the classical Boltzmann equation, 
with velocities equal to the group velocities of Bloch states.
%which are calculated by the first-principles method.
Boundary conditions for the distribution functions of electrons are determined by the scattering properties (transmission and reflection coefficients)
at the junction. % (scattering regions in this case).
SCB theory was implement previously by our group to study resistances of grain boundaries in metals, 
and well reproduced the experimental results\cite{PhysRevB.86.075134}.
In that work the transmission and reflection coefficients were computed
using a plane-wave based scattering formalism \cite{PhysRevB.70.045417},
which is computationally expensive for studying large systems.
%The remaining problem is that NEGF calculations do not provide transmission and reflection coefficients.
%To overcome this difficulty, we derived a generalized Fisher-Lee relation \cite{wimmer}. 
A feasible approach here is to use the NEGF technique implemented in a tight-binding framework
with an atomic orbital basis set, which allows us to treat a relatively large system
with sufficiently high numerical accuracy \cite{JChemPhys.128.114714}.  
Transmission and reflection coefficients can be calculated by the TB Fisher-Lee relation.

The remaining problem is to obtain group velocities from the NEGF calculations.
The group velocities of Bloch states in periodic leads 
corresponding to a given $\vec{k}_\|$
can be computed from the self-energies of the leads,
(the derivation is shown in Supplemental Materials),
\begin{equation}
\label{equ:groupv3}
v(k_z)=\frac{a_z}{\hbar}
u^\dagger_\lambda
\Gamma
u_\lambda
\end{equation}
where $\lambda=e^{ik_za_z}$,  $a_z$ is length of the lattice vector along the direction of transport, 
$\Gamma=i(\Sigma^r-\Sigma^a)$ where $\Sigma^{r/a}$ is the retarded/advanced self-energy of the leads,
and $ u_\lambda $ is the periodic part of Bloch state $\psi_\lambda(z) = e^{ik_z z} u_\lambda(z) $.
%Besides the group velocity, we also need to compute transmission and reflection coefficients from the Green's function.
In scattering theory, solutions of the Schr\"{o}dinger equation in the whole system
are constructed using the $r$ and $t$  coefficients, 
\begin{equation}
\Psi^M_{k_z}(\vec{r})  =\left\{
\begin{array}{cc}
\psi^M_{k_z,<}(\vec{r}) + \sum\limits_{N,k'_z} r_{k_z,k'_z}^{M} \psi^M_{k'_z,>}(\vec{r}) , & \hbox{$\vec{r}$ deep in $M$} \\
\noalign{\medskip}
\sum\limits_{N,k'_z} t_{k_z,k'_z}^{M,N} \psi^N_{k'_z,>}(\vec{r}) , & \hbox{$\vec{r}$ deep in  $N$}
\end{array}
\right.
\end{equation}
where outgoing(incoming) Bloch states are labelled by $>$($<$),
$M$ and $N$ are used for indices of leads connecting the scattering region.
In the TB Fisher-Lee relation, 
the transmission coefficients from a Bloch state in lead $M$ with $k_z$ to another state in a different lead $N$ with $k'_z$ 
are connected to Green's function in scattering region via\cite{wimmer}
\begin{equation}
\label{equ:t}
t^{M,N}_{k_z,k'_z} = \frac{i}
{\hbar\sqrt{|v_{<}^M(k_z)| \, |v_{>}^N(k'_z)|}}
u^{N,\dagger}_{k'_z,>}\Gamma_NG^r_{NM}\Gamma_Mu^{M}_{k_z,<}
\end{equation}
and the reflection via 
\begin{eqnarray}
\label{equ:r}
\nonumber
r^{M}_{k_z,k'_z} && = \frac{1}
{\hbar\sqrt{|v_{<}^M(k_z)| \,  |v_{>}^M(k'_z)|}} \times \\
&&\left[
iu^{M,\dagger}_{k'_z,>}\Gamma_MG^r_{MM}\Gamma_Mu^{M}_{k_z,<}
-u^{M,\dagger}_{k'_z,>}\Gamma_Mu^{M}_{k_z,<} \right] , 
\end{eqnarray}
where $G_{NM}$ and $G_{MM}$ are the corresponding sub-matrices of the Green's function matrix.
The energy dependences of $\Gamma$ and $G$ are omitted in the above equations.

We implemented Eqs.\eqref{equ:groupv3}--(\ref{equ:r})  in the
\texttt{TRANSIESTA} code \cite{PhysRevB.65.165401}, 
so that the group velocities of Bloch states in the leads as well as 
transmission and reflection coefficients of junctions at each $\vec{k}_\|$ can be computed
once the Green's function in the scattering region is obtained.
Finally the SCB equation was solved using $v$ [Eq.(\ref{equ:groupv3})], 
$t$ [Eq.(\ref{equ:t})], and $r$ [Eq.(\ref{equ:r})], as input.
In the CPP transport geometry, the resistance of the whole system
as the sum of the two leads and the junction
is obtained after solving the SCB equation.
The chemical potential along the direction of transport can also be obtained,
from which the resistance specifically of the junction can be computed.
An example to computed the resistance is shown in Supplemental Materials.

%\subsection{Results}

Table \ref{tab:RA} shows the products of resistance and area ($RA$) of Ag-silicene-Ag junctions obtained by solving the SCB equation.
Results from the Landauer-Buttiker equation are also shown for comparison.
It is clear that the Boltzmann transport formalism gives a lower resistance compared to the Landauer-Buttiker equation, 
since the latter neglects contributions to conductance from $\vec{k}_\|$.
Our results suggest that Ag-silicene-Ag junctions are well conducting, 
with resistances being from the same order up to several times larger
than those of grain boundaries in noble metals \cite{PhysRevB.86.075134}.

\begin{table}[t]
\begin{center}
\begin{ruledtabular}
\begin{tabular}{c c c}
Configuration & $RA$ (Boltzmann) & $RA$ (Landauer) \\
\hline
A $|$ 1Si $|$ A& 0.075 & 0.514 \\
A $|$ 1Si $|$ B& 0.085 & 0.512 \\
A $|$ 1Si $|$ C& 0.080 & 0.515 \\
\textbf{average}  & 0.080 & 0.514 \\
\hline
A $|$ 2Si $|$ A & 0.285 & 0.854 \\
A $|$ 2Si $|$ B & 0.316 & 0.877 \\
A $|$ 2Si $|$ C & 0.282 & 0.858 \\
\textbf{average} & 0.294 & 0.863\\
\end{tabular}
\end{ruledtabular}
\caption{
\label{tab:RA}
Products of resistance and area ($RA$) for Ag-silicene-Ag junctions in different configurations.
$RA$'s calculated from the semi-classical Boltzmann equation are in the middle column.
$RA$'s calculated from the Landauer-Buttiker equation are in the right-hand column for comparison.
All numbers are in units of $ \fom$.
}
\end{center}
%\vspace{-0.6cm}
\end{table}

%\section{Summary}
To summarize our findings, we conducted theoretical calculations to obtain the
interfacial structure of Ag-silicene-Ag junctions and study the interplay between
structure and transport characteristics.
Before silicene can be grown on or transferred onto insulating substrates, 
current-in-plane transport measurements do not reflect transport properties of standalone silicene.
We propose an experiment to study current-perpendicular-to-plane transport properties of silicene.
Our results show these junctions have relatively small resistances,
comparable to those of grain boundaries in metals.
The results indicate strong interaction between silicene and Ag,
in accord with our previous band structure study of silicene on Ag surfaces \cite{arxiv.1302.5759}.

To take the advantage of the efficient NEGF method and code, 
which makes it possible to deal with a large system, 
we developed a method that spans the NEGF method and scattering formalism, 
and which allows the group velocities of Bloch states in leads to be calculated from self-energies of leads and
transmission and reflection coefficients to be computed from the Green's function in the scattering region
via the TB Fisher-Lee relation.
With this development, we constructed a hybrid, two-scale framework to calculate resistance using the semi-classical Boltzmann equation
with input from the density-functional-based NEGF method.
Our scheme is suitable to study a suite of electron transport processes in 2D and 3D systems 
where a scattering potential
gives rise to transmission per channel greater than a typical tunneling junction.

\begin{acknowledgments}
This work was supported by the US Department of Energy (DOE), Office of Basic
Energy Sciences (BES), under Contract No. DE-FG02-02ER45995. Calculations 
were done using the utilities of the National Energy Research Scientific
Computing Center (NERSC).
\end{acknowledgments}

%%%%%%%%%%%%%%%%%%%%%%%%%%%%%%%%%%%%%%

%\bibliography{../../../../bib-collection}

\begin{thebibliography}{22}%
\makeatletter
\providecommand \@ifxundefined [1]{%
 \@ifx{#1\undefined}
}%
\providecommand \@ifnum [1]{%
 \ifnum #1\expandafter \@firstoftwo
 \else \expandafter \@secondoftwo
 \fi
}%
\providecommand \@ifx [1]{%
 \ifx #1\expandafter \@firstoftwo
 \else \expandafter \@secondoftwo
 \fi
}%
\providecommand \natexlab [1]{#1}%
\providecommand \enquote  [1]{``#1''}%
\providecommand \bibnamefont  [1]{#1}%
\providecommand \bibfnamefont [1]{#1}%
\providecommand \citenamefont [1]{#1}%
\providecommand \href@noop [0]{\@secondoftwo}%
\providecommand \href [0]{\begingroup \@sanitize@url \@href}%
\providecommand \@href[1]{\@@startlink{#1}\@@href}%
\providecommand \@@href[1]{\endgroup#1\@@endlink}%
\providecommand \@sanitize@url [0]{\catcode `\\12\catcode `\$12\catcode
  `\&12\catcode `\#12\catcode `\^12\catcode `\_12\catcode `\%12\relax}%
\providecommand \@@startlink[1]{}%
\providecommand \@@endlink[0]{}%
\providecommand \url  [0]{\begingroup\@sanitize@url \@url }%
\providecommand \@url [1]{\endgroup\@href {#1}{\urlprefix }}%
\providecommand \urlprefix  [0]{URL }%
\providecommand \Eprint [0]{\href }%
\providecommand \doibase [0]{http://dx.doi.org/}%
\providecommand \selectlanguage [0]{\@gobble}%
\providecommand \bibinfo  [0]{\@secondoftwo}%
\providecommand \bibfield  [0]{\@secondoftwo}%
\providecommand \translation [1]{[#1]}%
\providecommand \BibitemOpen [0]{}%
\providecommand \bibitemStop [0]{}%
\providecommand \bibitemNoStop [0]{.\EOS\space}%
\providecommand \EOS [0]{\spacefactor3000\relax}%
\providecommand \BibitemShut  [1]{\csname bibitem#1\endcsname}%
\let\auto@bib@innerbib\@empty
%</preamble>
\bibitem [{\citenamefont {Cahangirov}\ \emph {et~al.}(2009)\citenamefont
  {Cahangirov}, \citenamefont {Topsakal}, \citenamefont {Akt\"urk},
  \citenamefont {\ifmmode~\mbox{\c{S}}\else \c{S}\fi{}ahin},\ and\
  \citenamefont {Ciraci}}]{PhysRevLett.102.236804}%
  \BibitemOpen
  \bibfield  {author} {\bibinfo {author} {\bibfnamefont {S.}~\bibnamefont
  {Cahangirov}}, \bibinfo {author} {\bibfnamefont {M.}~\bibnamefont
  {Topsakal}}, \bibinfo {author} {\bibfnamefont {E.}~\bibnamefont {Akt\"urk}},
  \bibinfo {author} {\bibfnamefont {H.}~\bibnamefont
  {\ifmmode~\mbox{\c{S}}\else \c{S}\fi{}ahin}}, \ and\ \bibinfo {author}
  {\bibfnamefont {S.}~\bibnamefont {Ciraci}},\ }\href {\doibase
  10.1103/PhysRevLett.102.236804} {\bibfield  {journal} {\bibinfo  {journal}
  {Phys. Rev. Lett.}\ }\textbf {\bibinfo {volume} {102}},\ \bibinfo {pages}
  {236804} (\bibinfo {year} {2009})}\BibitemShut {NoStop}%
\bibitem [{\citenamefont {{Tsai Wei-Feng}}\ \emph {et~al.}(2013)\citenamefont
  {{Tsai Wei-Feng}}, \citenamefont {{Huang Cheng-Yi}}, \citenamefont {{Chang
  Tay-Rong}}, \citenamefont {{Lin Hsin}}, \citenamefont {{Jeng Horng-Tay}},\
  and\ \citenamefont {{Bansil A.}}}]{ncomms.4.1500}%
  \BibitemOpen
  \bibfield  {author} {\bibinfo {author} {\bibnamefont {{Tsai Wei-Feng}}},
  \bibinfo {author} {\bibnamefont {{Huang Cheng-Yi}}}, \bibinfo {author}
  {\bibnamefont {{Chang Tay-Rong}}}, \bibinfo {author} {\bibnamefont {{Lin
  Hsin}}}, \bibinfo {author} {\bibnamefont {{Jeng Horng-Tay}}}, \ and\ \bibinfo
  {author} {\bibnamefont {{Bansil A.}}},\ }\href {\doibase
  http://dx.doi.org/10.1038/ncomms2525} {\bibfield  {journal} {\bibinfo
  {journal} {Nat Commun}\ }\textbf {\bibinfo {volume} {4}},\ \bibinfo {pages}
  {1500} (\bibinfo {year} {2013})}\ \bibinfo {note}
  {}\BibitemShut {NoStop}%
\bibitem [{\citenamefont {Kang}\ \emph {et~al.}(2012)\citenamefont {Kang},
  \citenamefont {Wu},\ and\ \citenamefont {Li}}]{APL.100.233122}%
  \BibitemOpen
  \bibfield  {author} {\bibinfo {author} {\bibfnamefont {K.}~\bibnamefont
  {Kang}}, \bibinfo {author} {\bibfnamefont {F.}~\bibnamefont {Wu}}, \ and\
  \bibinfo {author} {\bibfnamefont {J.}~\bibnamefont {Li}},\ }\href@noop {}
  {\bibfield  {journal} {\bibinfo  {journal} {Appl. Phys. Lett.}\ }\textbf
  {\bibinfo {volume} {100}},\ \bibinfo {pages} {233122} (\bibinfo {year}
  {2012})}\BibitemShut {NoStop}%
\bibitem [{\citenamefont {Vogt}\ \emph {et~al.}(2012)\citenamefont {Vogt},
  \citenamefont {De~Padova}, \citenamefont {Quaresima}, \citenamefont {Avila},
  \citenamefont {Frantzeskakis}, \citenamefont {Asensio}, \citenamefont
  {Resta}, \citenamefont {Ealet},\ and\ \citenamefont
  {Le~Lay}}]{PhysRevLett.108.155501}%
  \BibitemOpen
  \bibfield  {author} {\bibinfo {author} {\bibfnamefont {P.}~\bibnamefont
  {Vogt}}, \bibinfo {author} {\bibfnamefont {P.}~\bibnamefont {De~Padova}},
  \bibinfo {author} {\bibfnamefont {C.}~\bibnamefont {Quaresima}}, \bibinfo
  {author} {\bibfnamefont {J.}~\bibnamefont {Avila}}, \bibinfo {author}
  {\bibfnamefont {E.}~\bibnamefont {Frantzeskakis}}, \bibinfo {author}
  {\bibfnamefont {M.~C.}\ \bibnamefont {Asensio}}, \bibinfo {author}
  {\bibfnamefont {A.}~\bibnamefont {Resta}}, \bibinfo {author} {\bibfnamefont
  {B.}~\bibnamefont {Ealet}}, \ and\ \bibinfo {author} {\bibfnamefont
  {G.}~\bibnamefont {Le~Lay}},\ }\href {\doibase
  10.1103/PhysRevLett.108.155501} {\bibfield  {journal} {\bibinfo  {journal}
  {Phys. Rev. Lett.}\ }\textbf {\bibinfo {volume} {108}},\ \bibinfo {pages}
  {155501} (\bibinfo {year} {2012})}\BibitemShut {NoStop}%
\bibitem [{\citenamefont {Chen}\ \emph {et~al.}(2012)\citenamefont {Chen},
  \citenamefont {Liu}, \citenamefont {Feng}, \citenamefont {He}, \citenamefont
  {Cheng}, \citenamefont {Ding}, \citenamefont {Meng}, \citenamefont {Yao},\
  and\ \citenamefont {Wu}}]{PhysRevLett.109.056804}%
  \BibitemOpen
  \bibfield  {author} {\bibinfo {author} {\bibfnamefont {L.}~\bibnamefont
  {Chen}}, \bibinfo {author} {\bibfnamefont {C.-C.}\ \bibnamefont {Liu}},
  \bibinfo {author} {\bibfnamefont {B.}~\bibnamefont {Feng}}, \bibinfo {author}
  {\bibfnamefont {X.}~\bibnamefont {He}}, \bibinfo {author} {\bibfnamefont
  {P.}~\bibnamefont {Cheng}}, \bibinfo {author} {\bibfnamefont
  {Z.}~\bibnamefont {Ding}}, \bibinfo {author} {\bibfnamefont {S.}~\bibnamefont
  {Meng}}, \bibinfo {author} {\bibfnamefont {Y.}~\bibnamefont {Yao}}, \ and\
  \bibinfo {author} {\bibfnamefont {K.}~\bibnamefont {Wu}},\ }\href {\doibase
  10.1103/PhysRevLett.109.056804} {\bibfield  {journal} {\bibinfo  {journal}
  {Phys. Rev. Lett.}\ }\textbf {\bibinfo {volume} {109}},\ \bibinfo {pages}
  {056804} (\bibinfo {year} {2012})}\BibitemShut {NoStop}%
\bibitem [{\citenamefont {Datta}(1995)}]{Datta1995}%
  \BibitemOpen
  \bibfield  {author} {\bibinfo {author} {\bibfnamefont {S.}~\bibnamefont
  {Datta}},\ }\href@noop {} {\emph {\bibinfo {title} {Electronic Transport in
  Mesoscopic Systems}}}\ (\bibinfo  {publisher} {Cambridge University Press},\
  \bibinfo {address} {Cambridge},\ \bibinfo {year} {1995})\BibitemShut
  {NoStop}%
\bibitem [{\citenamefont {Taylor}\ \emph {et~al.}(2001)\citenamefont {Taylor},
  \citenamefont {Guo},\ and\ \citenamefont {Wang}}]{PhysRevB.63.245407}%
  \BibitemOpen
  \bibfield  {author} {\bibinfo {author} {\bibfnamefont {J.}~\bibnamefont
  {Taylor}}, \bibinfo {author} {\bibfnamefont {H.}~\bibnamefont {Guo}}, \ and\
  \bibinfo {author} {\bibfnamefont {J.}~\bibnamefont {Wang}},\ }\href {\doibase
  10.1103/PhysRevB.63.245407} {\bibfield  {journal} {\bibinfo  {journal} {Phys.
  Rev. B}\ }\textbf {\bibinfo {volume} {63}},\ \bibinfo {pages} {245407}
  (\bibinfo {year} {2001})}\BibitemShut {NoStop}%
\bibitem [{\citenamefont {Xue}\ \emph {et~al.}(2002)\citenamefont {Xue},
  \citenamefont {Datta},\ and\ \citenamefont {Ratner}}]{Chem.Phys.281.151}%
  \BibitemOpen
  \bibfield  {author} {\bibinfo {author} {\bibfnamefont {Y.}~\bibnamefont
  {Xue}}, \bibinfo {author} {\bibfnamefont {S.}~\bibnamefont {Datta}}, \ and\
  \bibinfo {author} {\bibfnamefont {M.~A.}\ \bibnamefont {Ratner}},\
  }\href@noop {} {\bibfield  {journal} {\bibinfo  {journal} {Chem. Phys.}\
  }\textbf {\bibinfo {volume} {281}},\ \bibinfo {pages} {151} (\bibinfo {year}
  {2002})}\BibitemShut {NoStop}%
\bibitem [{\citenamefont {Fisher}\ and\ \citenamefont
  {Lee}(1981)}]{PhysRevB.23.6851}%
  \BibitemOpen
  \bibfield  {author} {\bibinfo {author} {\bibfnamefont {D.~S.}\ \bibnamefont
  {Fisher}}\ and\ \bibinfo {author} {\bibfnamefont {P.~A.}\ \bibnamefont
  {Lee}},\ }\href {\doibase 10.1103/PhysRevB.23.6851} {\bibfield  {journal}
  {\bibinfo  {journal} {Phys. Rev. B}\ }\textbf {\bibinfo {volume} {23}},\
  \bibinfo {pages} {6851} (\bibinfo {year} {1981})}\BibitemShut {NoStop}%
\bibitem [{\citenamefont {B\"uttiker}\ \emph {et~al.}(1985)\citenamefont
  {B\"uttiker}, \citenamefont {Imry}, \citenamefont {Landauer},\ and\
  \citenamefont {Pinhas}}]{PhysRevB.31.6207}%
  \BibitemOpen
  \bibfield  {author} {\bibinfo {author} {\bibfnamefont {M.}~\bibnamefont
  {B\"uttiker}}, \bibinfo {author} {\bibfnamefont {Y.}~\bibnamefont {Imry}},
  \bibinfo {author} {\bibfnamefont {R.}~\bibnamefont {Landauer}}, \ and\
  \bibinfo {author} {\bibfnamefont {S.}~\bibnamefont {Pinhas}},\ }\href
  {\doibase 10.1103/PhysRevB.31.6207} {\bibfield  {journal} {\bibinfo
  {journal} {Phys. Rev. B}\ }\textbf {\bibinfo {volume} {31}},\ \bibinfo
  {pages} {6207} (\bibinfo {year} {1985})}\BibitemShut {NoStop}%
\bibitem [{\citenamefont {Butler}\ \emph {et~al.}(2000)\citenamefont {Butler},
  \citenamefont {Zhang},\ and\ \citenamefont {MacLaren}}]{JS.13.221}%
  \BibitemOpen
  \bibfield  {author} {\bibinfo {author} {\bibfnamefont {W.~H.}\ \bibnamefont
  {Butler}}, \bibinfo {author} {\bibfnamefont {X.-G.}\ \bibnamefont {Zhang}}, \
  and\ \bibinfo {author} {\bibfnamefont {J.~M.}\ \bibnamefont {MacLaren}},\
  }\href@noop {} {\bibfield  {journal} {\bibinfo  {journal} {J. Superconduct.}\
  }\textbf {\bibinfo {volume} {13}},\ \bibinfo {pages} {221} (\bibinfo {year}
  {2000})}\BibitemShut {NoStop}%
\bibitem [{\citenamefont {Enriquez}\ \emph {et~al.}(2012)\citenamefont
  {Enriquez}, \citenamefont {Vizzini}, \citenamefont {Kara}, \citenamefont
  {Lalmi},\ and\ \citenamefont {Oughaddou}}]{0953-8984-24-31-314211}%
  \BibitemOpen
  \bibfield  {author} {\bibinfo {author} {\bibfnamefont {H.}~\bibnamefont
  {Enriquez}}, \bibinfo {author} {\bibfnamefont {S.}~\bibnamefont {Vizzini}},
  \bibinfo {author} {\bibfnamefont {A.}~\bibnamefont {Kara}}, \bibinfo {author}
  {\bibfnamefont {B.}~\bibnamefont {Lalmi}}, \ and\ \bibinfo {author}
  {\bibfnamefont {H.}~\bibnamefont {Oughaddou}},\ }\href
  {http://stacks.iop.org/0953-8984/24/i=31/a=314211} {\bibfield  {journal}
  {\bibinfo  {journal} {J. of Phys.: Condens. Matter}\ }\textbf {\bibinfo
  {volume} {24}},\ \bibinfo {pages} {314211} (\bibinfo {year}
  {2012})}\BibitemShut {NoStop}%
\bibitem [{\citenamefont {Perdew}\ \emph {et~al.}(1996)\citenamefont {Perdew},
  \citenamefont {Burke},\ and\ \citenamefont
  {Ernzerhof}}]{PhysRevLett.77.3865}%
  \BibitemOpen
  \bibfield  {author} {\bibinfo {author} {\bibfnamefont {J.~P.}\ \bibnamefont
  {Perdew}}, \bibinfo {author} {\bibfnamefont {K.}~\bibnamefont {Burke}}, \
  and\ \bibinfo {author} {\bibfnamefont {M.}~\bibnamefont {Ernzerhof}},\ }\href
  {\doibase 10.1103/PhysRevLett.77.3865} {\bibfield  {journal} {\bibinfo
  {journal} {Phys. Rev. Lett.}\ }\textbf {\bibinfo {volume} {77}},\ \bibinfo
  {pages} {3865} (\bibinfo {year} {1996})}\BibitemShut {NoStop}%
\bibitem [{\citenamefont {Kresse}\ and\ \citenamefont
  {Joubert}(1999)}]{PhysRevB.59.1758}%
  \BibitemOpen
  \bibfield  {author} {\bibinfo {author} {\bibfnamefont {G.}~\bibnamefont
  {Kresse}}\ and\ \bibinfo {author} {\bibfnamefont {D.}~\bibnamefont
  {Joubert}},\ }\href {\doibase 10.1103/PhysRevB.59.1758} {\bibfield  {journal}
  {\bibinfo  {journal} {Phys. Rev. B}\ }\textbf {\bibinfo {volume} {59}},\
  \bibinfo {pages} {1758} (\bibinfo {year} {1999})}\BibitemShut {NoStop}%
\bibitem [{\citenamefont {Lin}\ \emph {et~al.}(2013)\citenamefont {Lin},
  \citenamefont {Arafune}, \citenamefont {Kawahara}, \citenamefont {Kanno},
  \citenamefont {Tsukahara}, \citenamefont {Minamitani}, \citenamefont {Kim},
  \citenamefont {Kawai},\ and\ \citenamefont
  {Takagi}}]{PhysRevLett.110.076801}%
  \BibitemOpen
  \bibfield  {author} {\bibinfo {author} {\bibfnamefont {C.-L.}\ \bibnamefont
  {Lin}}, \bibinfo {author} {\bibfnamefont {R.}~\bibnamefont {Arafune}},
  \bibinfo {author} {\bibfnamefont {K.}~\bibnamefont {Kawahara}}, \bibinfo
  {author} {\bibfnamefont {M.}~\bibnamefont {Kanno}}, \bibinfo {author}
  {\bibfnamefont {N.}~\bibnamefont {Tsukahara}}, \bibinfo {author}
  {\bibfnamefont {E.}~\bibnamefont {Minamitani}}, \bibinfo {author}
  {\bibfnamefont {Y.}~\bibnamefont {Kim}}, \bibinfo {author} {\bibfnamefont
  {M.}~\bibnamefont {Kawai}}, \ and\ \bibinfo {author} {\bibfnamefont
  {N.}~\bibnamefont {Takagi}},\ }\href {\doibase
  10.1103/PhysRevLett.110.076801} {\bibfield  {journal} {\bibinfo  {journal}
  {Phys. Rev. Lett.}\ }\textbf {\bibinfo {volume} {110}},\ \bibinfo {pages}
  {076801} (\bibinfo {year} {2013})}\BibitemShut {NoStop}%
\bibitem [{\citenamefont {Troullier}\ and\ \citenamefont
  {Martins}(1991)}]{PhysRevB.43.1993}%
  \BibitemOpen
  \bibfield  {author} {\bibinfo {author} {\bibfnamefont {N.}~\bibnamefont
  {Troullier}}\ and\ \bibinfo {author} {\bibfnamefont {J.~L.}\ \bibnamefont
  {Martins}},\ }\href {\doibase 10.1103/PhysRevB.43.1993} {\bibfield  {journal}
  {\bibinfo  {journal} {Phys. Rev. B}\ }\textbf {\bibinfo {volume} {43}},\
  \bibinfo {pages} {1993} (\bibinfo {year} {1991})}\BibitemShut {NoStop}%
\bibitem [{\citenamefont {Brandbyge}\ \emph {et~al.}(2002)\citenamefont
  {Brandbyge}, \citenamefont {Mozos}, \citenamefont {Ordej\'on}, \citenamefont
  {Taylor},\ and\ \citenamefont {Stokbro}}]{PhysRevB.65.165401}%
  \BibitemOpen
  \bibfield  {author} {\bibinfo {author} {\bibfnamefont {M.}~\bibnamefont
  {Brandbyge}}, \bibinfo {author} {\bibfnamefont {J.-L.}\ \bibnamefont
  {Mozos}}, \bibinfo {author} {\bibfnamefont {P.}~\bibnamefont {Ordej\'on}},
  \bibinfo {author} {\bibfnamefont {J.}~\bibnamefont {Taylor}}, \ and\ \bibinfo
  {author} {\bibfnamefont {K.}~\bibnamefont {Stokbro}},\ }\href {\doibase
  10.1103/PhysRevB.65.165401} {\bibfield  {journal} {\bibinfo  {journal} {Phys.
  Rev. B}\ }\textbf {\bibinfo {volume} {65}},\ \bibinfo {pages} {165401}
  (\bibinfo {year} {2002})}\BibitemShut {NoStop}%
\bibitem [{\citenamefont {Srivastava}\ \emph {et~al.}(2012)\citenamefont
  {Srivastava}, \citenamefont {Wang}, \citenamefont {Zhang}, \citenamefont
  {Nicholson},\ and\ \citenamefont {Cheng}}]{PhysRevB.86.075134}%
  \BibitemOpen
  \bibfield  {author} {\bibinfo {author} {\bibfnamefont {M.~K.}\ \bibnamefont
  {Srivastava}}, \bibinfo {author} {\bibfnamefont {Y.}~\bibnamefont {Wang}},
  \bibinfo {author} {\bibfnamefont {X.-G.}\ \bibnamefont {Zhang}}, \bibinfo
  {author} {\bibfnamefont {D.~M.~C.}\ \bibnamefont {Nicholson}}, \ and\
  \bibinfo {author} {\bibfnamefont {H.-P.}\ \bibnamefont {Cheng}},\ }\href
  {\doibase 10.1103/PhysRevB.86.075134} {\bibfield  {journal} {\bibinfo
  {journal} {Phys. Rev. B}\ }\textbf {\bibinfo {volume} {86}},\ \bibinfo
  {pages} {075134} (\bibinfo {year} {2012})}\BibitemShut {NoStop}%
\bibitem [{\citenamefont {Smogunov}\ \emph {et~al.}(2004)\citenamefont
  {Smogunov}, \citenamefont {Dal~Corso},\ and\ \citenamefont
  {Tosatti}}]{PhysRevB.70.045417}%
  \BibitemOpen
  \bibfield  {author} {\bibinfo {author} {\bibfnamefont {A.}~\bibnamefont
  {Smogunov}}, \bibinfo {author} {\bibfnamefont {A.}~\bibnamefont {Dal~Corso}},
  \ and\ \bibinfo {author} {\bibfnamefont {E.}~\bibnamefont {Tosatti}},\ }\href
  {\doibase 10.1103/PhysRevB.70.045417} {\bibfield  {journal} {\bibinfo
  {journal} {Phys. Rev. B}\ }\textbf {\bibinfo {volume} {70}},\ \bibinfo
  {pages} {045417} (\bibinfo {year} {2004})}\BibitemShut {NoStop}%
\bibitem [{\citenamefont {Strange}\ \emph {et~al.}(2008)\citenamefont
  {Strange}, \citenamefont {Kristensen}, \citenamefont {Thygesen},\ and\
  \citenamefont {Jacobsen}}]{JChemPhys.128.114714}%
  \BibitemOpen
  \bibfield  {author} {\bibinfo {author} {\bibfnamefont {M.}~\bibnamefont
  {Strange}}, \bibinfo {author} {\bibfnamefont {I.~S.}\ \bibnamefont
  {Kristensen}}, \bibinfo {author} {\bibfnamefont {K.~S.}\ \bibnamefont
  {Thygesen}}, \ and\ \bibinfo {author} {\bibfnamefont {K.~W.}\ \bibnamefont
  {Jacobsen}},\ }\href {\doibase http://dx.doi.org/10.1063/1.2839275}
  {\bibfield  {journal} {\bibinfo  {journal} {J. Chem. Phys}\ }\textbf
  {\bibinfo {volume} {128}},\ \bibinfo {pages} {114714} (\bibinfo {year}
  {2008})}\BibitemShut {NoStop}%
\bibitem [{\citenamefont {Wimmer}(2009)}]{wimmer}%
  \BibitemOpen
  \bibfield  {author} {\bibinfo {author} {\bibfnamefont {M.}~\bibnamefont
  {Wimmer}},\ }\href@noop {} {\emph {\bibinfo {title} {Quantum transport in
  nanostructures: from computational concepts to spintronics in graphene and
  magnetic tunnel junctions}}}\ (\bibinfo  {publisher} {Dover},\ \bibinfo
  {year} {2009})\BibitemShut {NoStop}%
\bibitem [{\citenamefont {Wang}\ and\ \citenamefont
  {Cheng}(2013)}]{arxiv.1302.5759}%
  \BibitemOpen
  \bibfield  {author} {\bibinfo {author} {\bibfnamefont {Y.-P.}\ \bibnamefont
  {Wang}}\ and\ \bibinfo {author} {\bibfnamefont {H.-P.}\ \bibnamefont
  {Cheng}},\ }\href@noop {} {\  (\bibinfo {year} {2013})},\ \Eprint
  {http://arxiv.org/abs/arXiv:1302.5759} {arXiv:1302.5759} \BibitemShut
  {NoStop}%
\end{thebibliography}


%%%%%%%%%%%%%%%%%%%%%%%%%%%%%%%%%%%%%%

\end{document}